\documentclass[fleqn,usenatbib]{mnras}

\usepackage{newtxtext,newtxmath}

\usepackage[T1]{fontenc}

\DeclareRobustCommand{\VAN}[3]{#2}
\let\VANthebibliography\thebibliography
\def\thebibliography{\DeclareRobustCommand{\VAN}[3]{##3}\VANthebibliography}

\usepackage{graphicx}	
\usepackage{amsmath}	
\usepackage{soul}

\title[Diffused GRBs from AGNs]{The emergence of diffused Gamma-Ray Burst afterglows from the disks of Active Galactic Nuclei }

\author[Wang et al.]{
Yi-Han Wang$^{1}$\thanks{E-mail:yihan.wang.1@stonybrook.edu },
Davide Lazzati$^{2}$,
Rosalba Perna$^{1,3}$
\\
$^{1}$ Department of Physics and Astronomy, Stony Brook
  University, Stony Brook, NY 11794-3800, USA\\
$^{2}$ Department of Physics, Oregon State University, 301
  Weniger Hall, Corvallis, OR 97331, USA\\
$^{3}$ Center for Computational Astrophysics, Flatiron Institute, New York, NY 10010, USA
}

\date{Accepted XXX. Received YYY; in original form ZZZ}

\pubyear{2015}

\begin{document}
\label{firstpage}
\pagerange{\pageref{firstpage}--\pageref{lastpage}}
\maketitle

\begin{abstract}
{The disks of Active Galactic Nuclei (AGNs) have emerged as rich
environments for the production and capture of stars and the compact
objects that they leave behind. These stars  produce long Gamma-Ray Bursts (LGRBs) at their
deaths, while frequent interactions among compact objects form binary neutron stars and neutron
star-black hole binaries, leading to short GRBs (SGRBs) upon their
merger. Predicting the properties of these transients as they emerge from the dense
environments of	AGN disks is key to their proper identification and 
to better constrain the	star and compact object	population in AGN disks. Some of these transients would appear unusual because they take place in much higher densities than the interstellar medium. Others, which are the subject of this paper, would additionally be modified by radiation diffusion, since they are generated within optically thick regions of the accretion disks. 
Here we compute	the GRB afterglow light curves for diffused GRB sources for a
representative variety of central black-hole masses and	disk locations.
We find that the radiation from radio to UV and soft X-rays can be strongly suppressed by synchrotron self-absorption in the dense medium of the AGN disk. In addition, photon diffusion can significantly delay the emergence of the emission peak, turning a beamed, fast transient into a slow, isotropic, and dimmer one. 
These  would appear as broadband-correlated  AGN variability with a dominance at the higher frequencies. Their properties can constrain both the stellar populations within AGN disks as well as the disk structure.}

\end{abstract}

\begin{keywords}
accretion, accretion disks -- galaxies: active -- gamma-ray burst: general
\end{keywords}



\section{Introduction}

The disks of Active Galactic Nuclei (AGNs) have recently emerged as a potentially interesting site for hosting some of the most unexpected LIGO/Virgo binary black hole (BH)  merger events \citep{Perna2021}, such as BHs in the low mass gap \citep{Abbott2020low}, as well as BHs with masses above what generally believed to be the gap due to the onset of the pair instability \citep{Abbott2020high}. While traditional channels cannot be excluded \citep{Farmer2020,Belczynski2020}, AGN disks provide a natural environment for BH growth, due to both accretion from the very dense medium \citep{Yang2020,Tagawa2020}, as well as via hierarchical mergers \citep{McKernan2020,Tagawa2020,Li2022}. Other unusual properties, such as an observed asymmetry in the spin distribution \citep{Callister2021} can similarly be accounted via the AGN disk host channel \citep{McKernan2021,Wang2021}. 

The presence of black holes, and more generally compact objects in AGN disks, is not surprising, since stars are believed to exist in these environments. Both in-situ formation triggered by gravitational instabilities in the outer parts of the disk \citep{Goodman2003, Dittmann2020}, as well as capture from the nuclear star cluster surrounding the AGN disk \citep{Artymowicz1993}, contribute to this population. 

The evolution of stars in the specific environment of an AGN disk has been the subject of recent investigations
\citep{Cantiello2021,Dittmann2021,Jermyn2021}. These have uncovered peculiar features due to accretion from the AGN disk, such as growth to masses $\gtrsim 100 M_\odot$, as well as high rotational speeds of the pre-supernova star. These conditions are conducive to these massive stars to produce long Gamma-Ray Bursts (GRBs) accompanying the supernova explosions \citep{MacFadyen1999}. { Furthermore, mergers between neutron stars (NSs) and possibly a fraction of the NS-BH systems in the AGN disk will give rise to short GRBs.}   

\begin{figure*}
\includegraphics[width=2\columnwidth]{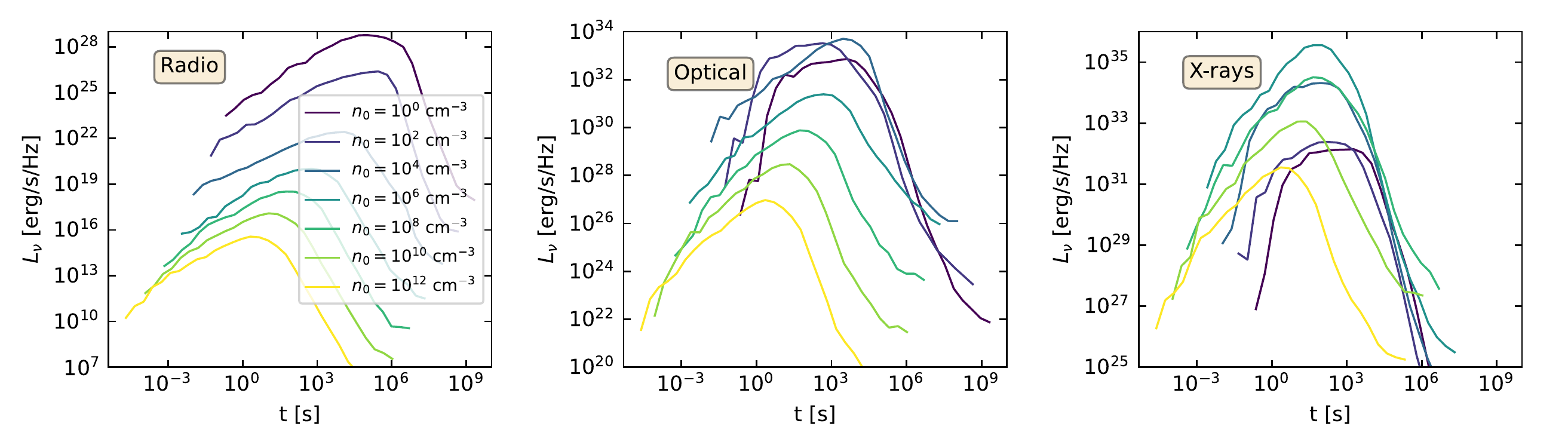}
\caption{ Face-on (5$^\circ$ viewing angle) afterglow light curves in three representative bands: Radio (left), Optical (middle) and X-ray (right) for a wide range of densities, from the one typical of the interstellar medium to the high range typical of the disks of Active Galactic Nuclei. }
\label{fig:lightcurves}
\end{figure*}

\begin{figure*}
\includegraphics[width=\columnwidth]{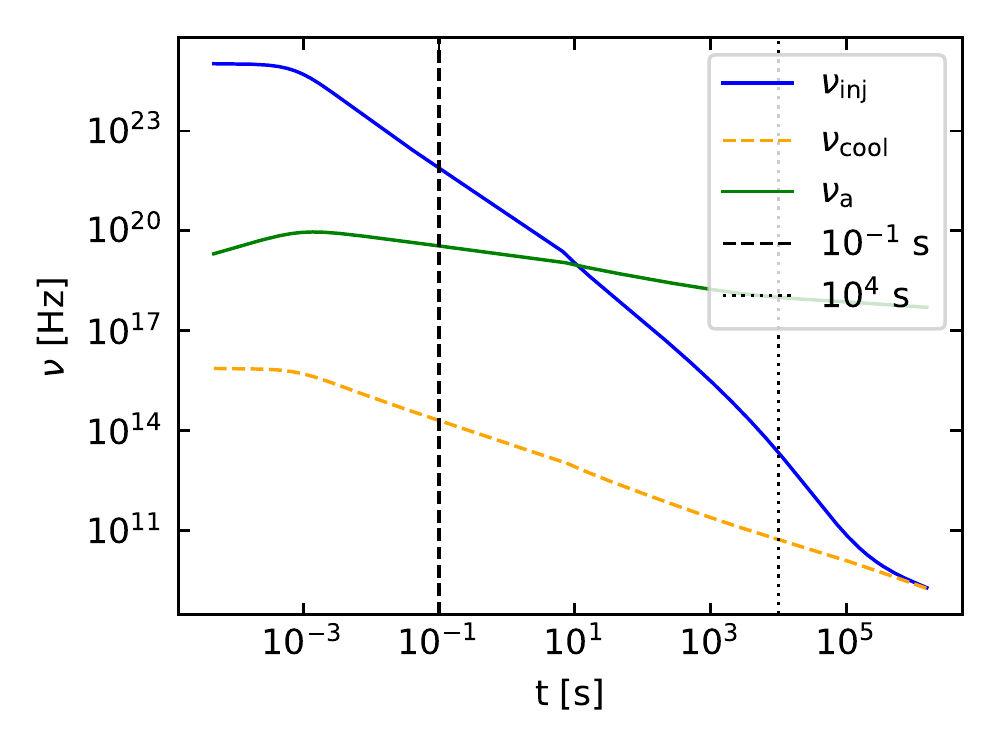}
\includegraphics[width=\columnwidth]{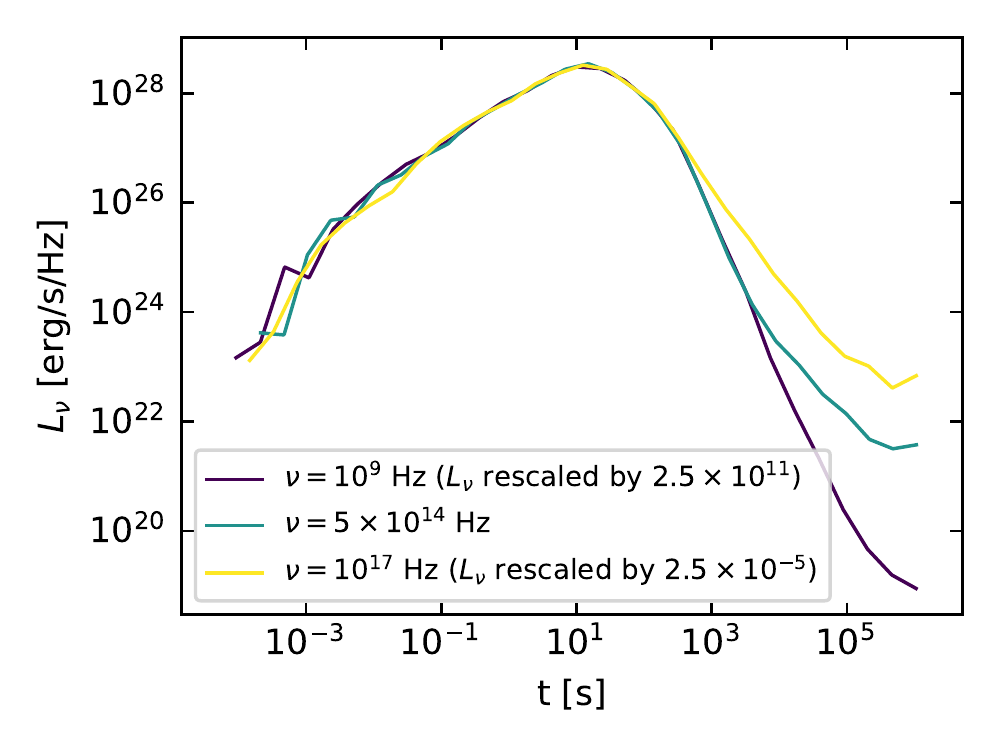}\\
\includegraphics[width=\columnwidth]{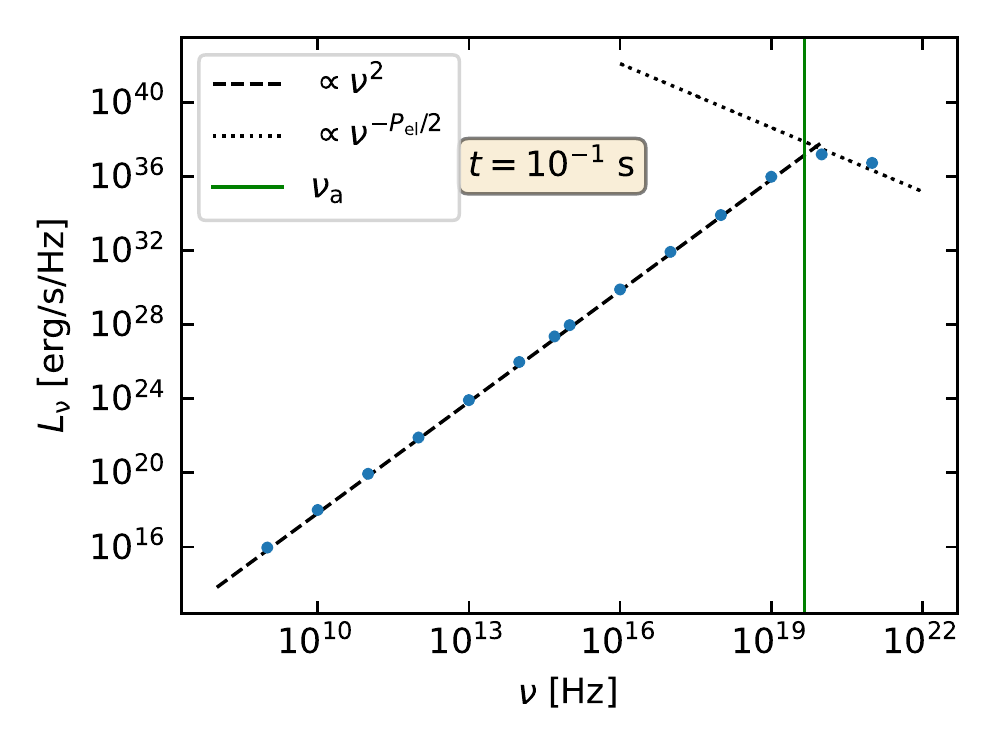}
\includegraphics[width=\columnwidth]{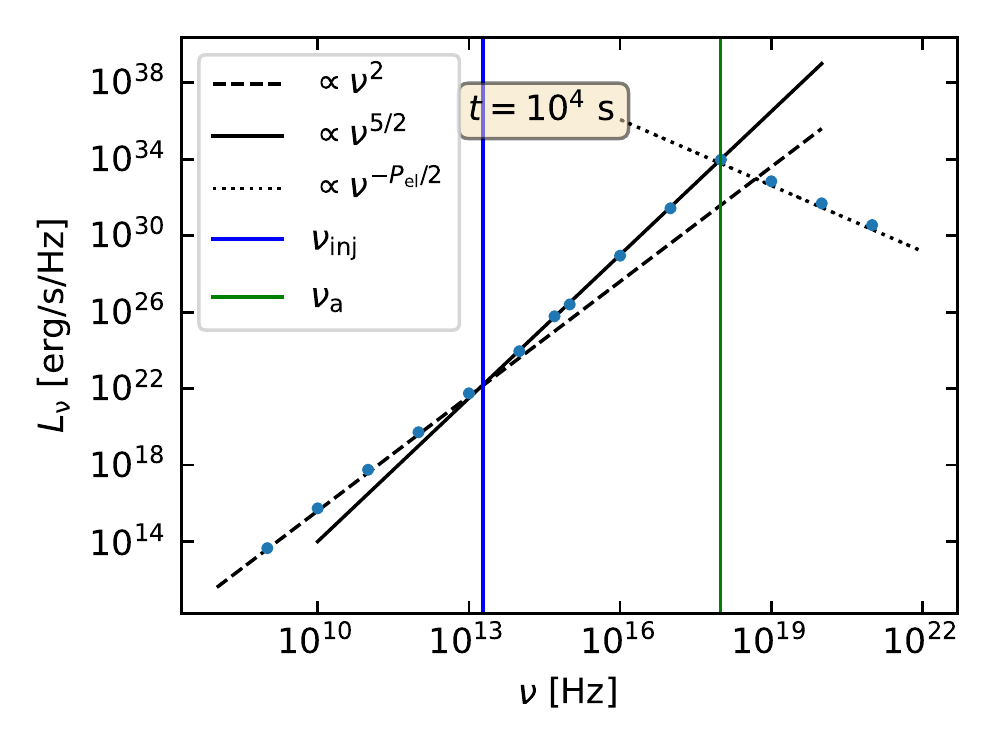}
\caption{\textit{Upper left} panel:  Injected, cooling and absorption frequency of a GRB afterglow as a function of time. The afterglow model is generated in a  medium of density $n_0 = 10^{10}$ cm$^{-3}$; the source has a Lorentz factor $\Gamma_0=300$ and an outflow energy $E_{\rm iso}=10^{53}$ erg. \textit{Upper right} panel: Face-on light curve ($5^\circ$ viewing angle) of the afterglow in the Radio ($10^{9}$ Hz), Optical ($5\times10^{14}$ Hz) and X-ray ($10^{17}$ Hz) bands. \textit{Bottom left} and \textit{bottom right} panels:  The spectrum of the afterglow at two representative times.}
\label{fig:spectrum}
\end{figure*}

In addition to the long and short GRB transients, other transients may also find a conducive environment to their production in AGN disks, and in particular tidal disruption events by stellar-mass black holes \citep{Yang2021}, accretion-induced collapse of NSs \citep{Perna2021b,Pan2021} and White Dwarfs \citep{Zhu2021WD}, jets from accreting BHs \citep{Tagawa2022} and binary BHs \citep{Wang2021EM}, super-kilonovae \citep{Siegel2021}, BH-BH mergers \citep{Bartos2017,Graham2020,Kaaz2021},
as well as supernovae \citep{Grishin2021}. 

The identification of electromagnetic transients (EM) emerging from AGN disks is an exciting prospect, as it would help calibrate the star and compact object populations in these disks, as well as learn more about some of the exotic phenomena specific to these environments. However, the extremely high densities which characterize AGN disks can significantly modify the emerging { light curves and } spectra, hence hampering our ability to recognize and identify transients based on our knowledge from their appearance in more typical galactic environments.
In \citet{Perna2021} we began to investigate the general question of the observability of relativistic transients from AGN disks (see also \citealt{Zhu2021a}). The analysis revealed that the extent to which the transient is modified is very sensitive to its location within the disk, being especially influenced by whether the (time-dependent) emission from the moving jet happens when the jet is below or above the disk photosphere. {That} work identified the various timescales on which the transients are expected to be seen by the observer, depending on their initial location within the disk. 

Here we continue our investigation of the emergence of relativistic transients with a radiative transfer calculation of light curves and spectra, focusing on transients whose afterglow is at least partially diffused by the AGN material. We note that our analysis for the propagation of photons within an AGN disk  { applies to all transients generated by relativistically expanding material (e.g., Tidal Disruption Events)}, even though this paper focuses on the input spectrum of a GRB afterglow (since these are well studied transients).
The paper is organized as follows: Sec.~2 describes the input afterglow spectrum and the numerical methods used to compute its propagation within the AGN disk. Results for a variety of locations within the disk are displayed and discussed in Sec.3. We summarize and conclude in Sec.4.

\begin{figure}
\includegraphics[width=\columnwidth]{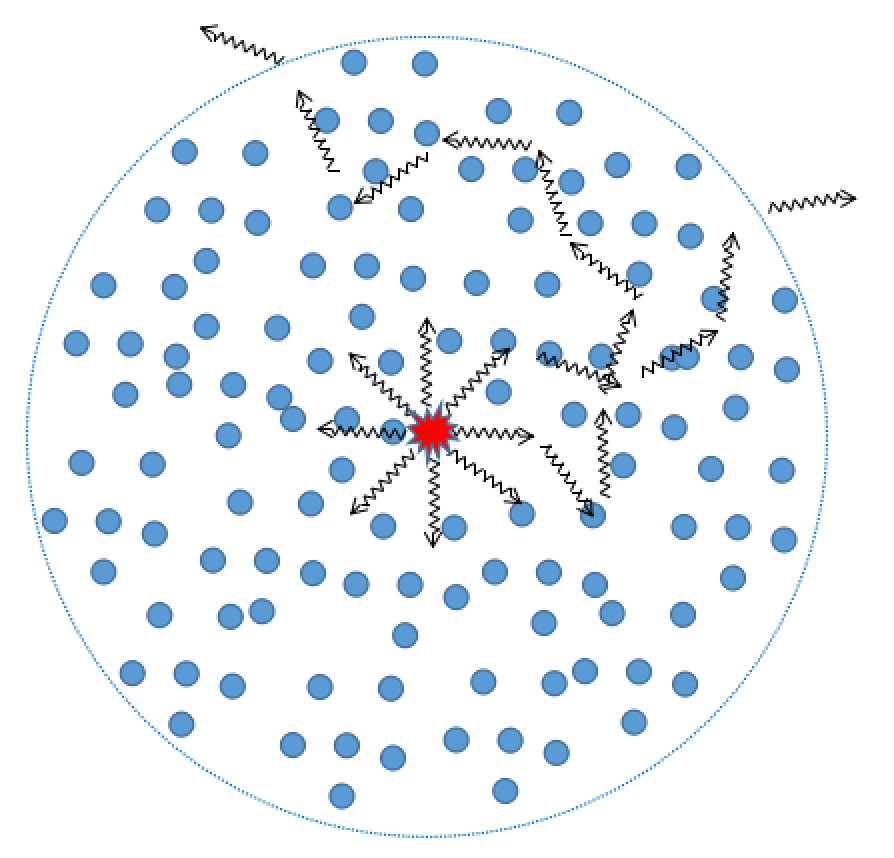}
\caption{Schematics of the toy model, where an isotropic, pulse-like source is placed at the center of a spherical region with uniform medium density. The radius of the region is assumed to be $10^{16}$~cm,  and the number density of the medium is varied in a range of values between  $10^8$  and $5\times 10^{10}~\rm cm^{-3}$.}
\label{fig:toy-schematics}
\end{figure}

\section{Numerical Methods: Afterglow production and propagation within the AGN disk}

\subsection{GRB afterglow model: general}\label{sec:afterglow-gen}

We use a line of sight afterglow model to calculate the intrinsic afterglow light curves, 
before the diffusion process through the disk material. { Line of sight} models do not include details such as the equal arrival time surfaces of photons \citep{Panaitescu1998} or the thickness of the blastwave \citep{Blandford1976} but are fast and reliable. Considering that all short time-scale features are washed out by the diffusion, we consider a line of sight algorithm sufficient for the scope of this paper.

\begin{figure*}
\includegraphics[width=2\columnwidth]{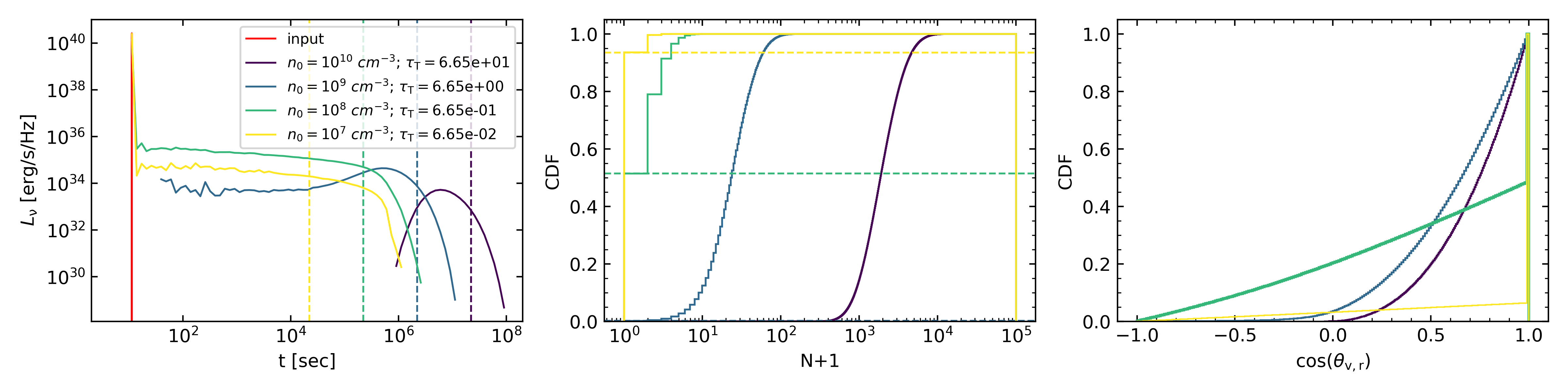}
\caption{{\em Left:}  The emergent, diffused light curve for the toy model illustrated in Fig.~\ref{fig:toy-schematics}. 
The input light curve is modeled as a pulse at $t=10$~s (for convenience in the log-plot)
with total energy $10^{53}$~erg.
The scattering region has radius of $10^{16}$~cm and variable density as indicated in the label. {\em Middle:} The corresponding  number of scatterings for each of the photons as they emerge from 
the scattering regions with densities as
in the left panel. The fraction of photons which is unscattered (ie. with $N=0$) in the lower density  cases emerges in correspondence to the time of the input pulse. The horizontal dashed lines show the transmittance of the medium, matching the fraction of unscattered photons. {\em Right:} The distribution of the angle between the velocity vector and the location vector of the scattered photons.}
\label{fig:toy}
\end{figure*}

The fireball dynamics is calculated by assuming energy conservation, yielding (e.g., \citealt{Paczynski1993})
\begin{equation}
    \Gamma(r)=\frac{\Gamma_0+f(r)}{\sqrt{1+2\Gamma_0 f(r) + f^2(r)}}
\end{equation}
where { $r$ is the radial distance to the burst engine,} $\Gamma_0$ is the initial Lorentz factor of the fireball and $f(r)$ is the ratio of the ISM mass swept up by the fireball over the  initial rest mass of the fireball.

A constant fraction $\epsilon_e$ of the internal blastwave energy is used to accelerate electrons in a non-thermal power-law energy distribution with a typical Lorentz factor 
\begin{equation}
    \gamma_i=\epsilon_e (\Gamma(r)-1) \frac{m_p}{m_e}
\end{equation}
and power-law slope $p_{\rm{el}}$, where $m_p$ is the proton mass and $m_e$ the electron mass. A constant fraction $\epsilon_B$ of the internal blastwave energy is instead used to generate a randomly oriented magnetic field with intensity
\begin{equation}
    \frac{B^2{ (r)}}{8\pi}=4\epsilon_B m_p c^2 n_0\left(\Gamma(r)-1\right)\left(\Gamma(r)+\frac34\right)\,,
\end{equation}
where $n_0$ is the density of the external medium \citep{Panaitescu2000}.

In these conditions the fireball radiates via the synchrotron mechanism, with a spectrum characterized by a peak  specific luminosity
\begin{equation}
    L_{\nu_{\rm{pk}}}{ (r)}=\frac{e^3}{m_e c^2}\Gamma(r) B N_e\,,
    \label{eq:paekL}
\end{equation}
where $N_e$ is the total number of ambient electrons swept up by the fireball \citep{Panaitescu2000}. The radiated spectrum is also characterized by spectral breaks at the self-absorption frequency and at other frequencies that correspond to significant electron energies: the energy of electrons that are newly injected in the blastwave and the energy of electrons that cool radiatively \citep{Sari1998}. The determination of the break frequency and the spectral shape for the specific case of high density environments is described below.

\subsection{GRB afterglow model in dense media}\label{sec:GRB-afterglow}

In the following we describe how we compute light curves for afterglows in 
high density media. The unique challenge of this step is the need to consider cases in which the self-absorption frequency is the largest among all the break frequencies (the other two being the injection frequency and the cooling frequency). This unusual ordering of the breaks is due to the fact that the self-absorption frequency grows with external density (e.g., Eq. 55 and 60 in \citealt{Panaitescu2000}). The injection frequency, on the other hand, is independent of the external density, while the cooling frequency is inversely proportional to the external density (e.g., Eq. 22 and 27 in \citealt{Panaitescu2000}). For this reason, the most likely case in high density external media is one in which the self absorption frequency is in the X-ray band or above. It is worth nothing here that whenever the cooling frequency is formally smaller than the self-absorption frequency, the cooling frequency disappears, since electrons in the self-absorption regime do not cool.

To better understand the shape of the spectrum in this circumstance, let us consider the electron population. Electrons are injected at the forward shock with a Lorentz factor distribution characterized by a power-law of slope $p_{\rm{el}}$ and a minimum Lorentz factor $\gamma_{\rm{inj}}$. Electrons with a Lorentz factor that is larger than  {  the value  $\gamma_{a}$
at which self-absorption becomes dominant} are in the fast cooling regime, since the formal cooling frequency is smaller than the self-absorption one. Electrons with $\gamma\le\gamma_a$ are instead unable to cool because they are in self-absorption regime. A steady state electron distribution is therefore established as
\begin{equation}
    \frac{dN}{d\gamma}\propto
    \left\{
    \begin{array}{ll}
    \gamma^{-p_{\rm{el}}} & \gamma_{\rm{inj}}\le\gamma\le\gamma_a \\
    \gamma^{-(p_{\rm{el}}+1)} & \gamma>\gamma_a
    \end{array}
    \right.
\end{equation}
which, in turn, results in a spectrum:
\begin{equation}
    L(\nu) = L(\nu_a) \times
    \left\{
    \begin{array}{ll}
    \left(\frac{\nu_{\rm{inj}}}{\nu_a}\right)^{\frac52}\left(\frac{\nu}{\nu_{\rm{inj}}}\right)^{2} & \nu<\nu_{\rm{inj}} \\
    \left(\frac{\nu}{\nu_a}\right)^{\frac52} & \nu_{\rm{inj}}\le\nu\le\nu_a \\
    \left(\frac{\nu}{\nu_a}\right)^{-\frac{p_{\rm{el}}}{2}} & \nu>\nu_a \end{array}
    \right.
\end{equation}
where $\nu_{\rm{inj}}$ and $\nu_a$ are the injection and self-absorption frequencies, respectively, and $L(\nu_a)$ is the flux at the peak of the spectrum given by Equation~\ref{eq:paekL} (see also Eq. 64 of \citealt{Panaitescu2000}). 

In order to illustrate the changing behaviour of the light curves with ambient density, we show in Figure~\ref{fig:lightcurves} several examples in three representative bands (radio, optical and X-rays)
as the medium density increases from that of the typical galactic interstellar medium (ISM) $n_0\sim 1$~cm$^{-3}$ to a high value as in the interior of an AGN disk, $n_0\sim 10^{12}$~cm$^{-3}$. The engine is assumed to be a jet of isotropic energy $E_{\rm iso}=10^{53}$~erg, Lorentz factor $\Gamma_0=300$ and opening angle $\theta_{\rm jet}=15^\circ$. 
We note that the radio, due to the $\nu^2$ cutoff resulting from self-absorption, is highly suppressed compared to what it would be from a similar source in the interstellar medium. For medium densities of $n_0\sim 10^{12}$~cm$^{-3}$, the peak flux is about 15 orders of magnitude lower than for a typical ISM of $n_0\sim 1$~cm$^{-3}$. Note that the flux also peaks at earlier times due to the fact that the jet break time
\begin{equation}
t_{\rm break} \approx 20.6 
\left(\frac{E_{\rm iso}}{10^{53}~{\rm erg}}\right)^{1/3}
\left(\frac{n_0}{{\rm cm}^{-3}}\right)^{-1/3}
\left(\frac{\theta_{\rm jet}}{0.1~{\rm rad}}\right)~{\rm hr}
\label{eq:jbreak}    
\end{equation}
occurs at earlier times at higher densities \citep{Sari1999}.
A large flux suppression is also observed in the optical, albeit to a lesser extent than in the radio, with a peak flux reduced by about 8 orders of magnitudes between the same density range quoted above.  
Generally, at frequencies which are usually unabsorbed, the flux initially increases with density. However, when the absorption frequency crosses the observing band, the flux starts to become suppressed. This happens at different densities for different observing bands.
In particular, in the optical band 
it occurs  for densities of $n_0\sim 10^2$~cm$^{-3}$ while in the X-rays for 
$n_0\sim 10^{6}$~cm$^{-3}$ .

An example of spectrum and light curve are shown in Figure~\ref{fig:spectrum} for a medium density of $10^{10}$~cm$^{-3}$ as typical of some regions in the AGN disk
and same engine parameters. 
More specifically, the upper left panel shows the injected, cooling and absorption frequency $\nu_{\rm inj},\nu_{\rm c}$ and $\nu_{\rm a}$ respectively as a function of time. The upper right panel shows the light curve at three representative frequencies in the X-ray, optical, and radio bands, while the two bottom panels display the spectrum at two representative times.

\subsection{AGN Disk Model}\label{sec:disk}
We adopt the Thompson et al. (TQM) model as our fiducial AGN disk model, and assume that the density profile in the $z$ direction (perpendicular to the disk plane) can be described as
\begin{equation}
    \rho_{\rm disk}(r,z) = \rho(r)\exp^{-\frac{z^2}{2h^2(r)}}\,,
\end{equation}
where $\rho(r)$ and $h(r)$ are central density and scale height given by TQM model. A graphic representation of the density and scale height profiles with radius for this model can be seen, e.g., in Fig.~1 of \citet{Fabj2020}.

\subsection{Monte Carlo radiative transfer code} 
The {  rest-mass energy of an} electron is $m_ec^2\sim 511$ keV with corresponding frequency $\sim$ 1.2$\times10^{20}$ Hz. If the energy of a scattering photons is much smaller than the rest energy of the electrons, which is true for all wavebands up to soft X-rays, the scatterings between the photons and the electrons are elastic {  to a very good approximation}, resulting in no energy shift for the photons from the scatterings. Hence Thompson scattering is a good approximation for our study.
 
{  In the following we describe
how we follow photon propagation in a dense environment, from the moment the photons
are produced at the source, to the time they reach the photosphere of the absorbing region. We developed a  Monte Carlo code to follow the individual trajectories of the photons as }
 they are scattered by ionized electrons { in} the disk\footnote{We note that, even though the external regions of the disk may initially be neutral, the early X-ray/UV radiation from the burst itself will ionize it on a very short timescale \citep{Perna2021}.}.
 { 
 We generate the initial positions and velocity vectors of photons at each frequency and time in the lab frame based on the source model parameters
 which, for GRB afterglows, are described in Section~\ref{sec:afterglow-gen} and \ref{sec:GRB-afterglow}.} {  A challenge is represented by the fact that the input light curve spans a wide range of luminosities. To ensure similar accuracy for bright and dim segments of the light curve,  we assign a weight to the generated photons. With this procedure we assign a larger weight to the photons in bright segments and a smaller weight to photons in dim segments, maintaining a fairly constant number of photons throughout the entire light curve}. {  The weights are normalized such that the total energy of the weighted photons is consistent with the afterglow input light curve (energy conservation). For each input afterglow light curve we generate $10^6$ weighted photons to be used in the diffusion code.
}

\begin{figure}
\includegraphics[width=\columnwidth]{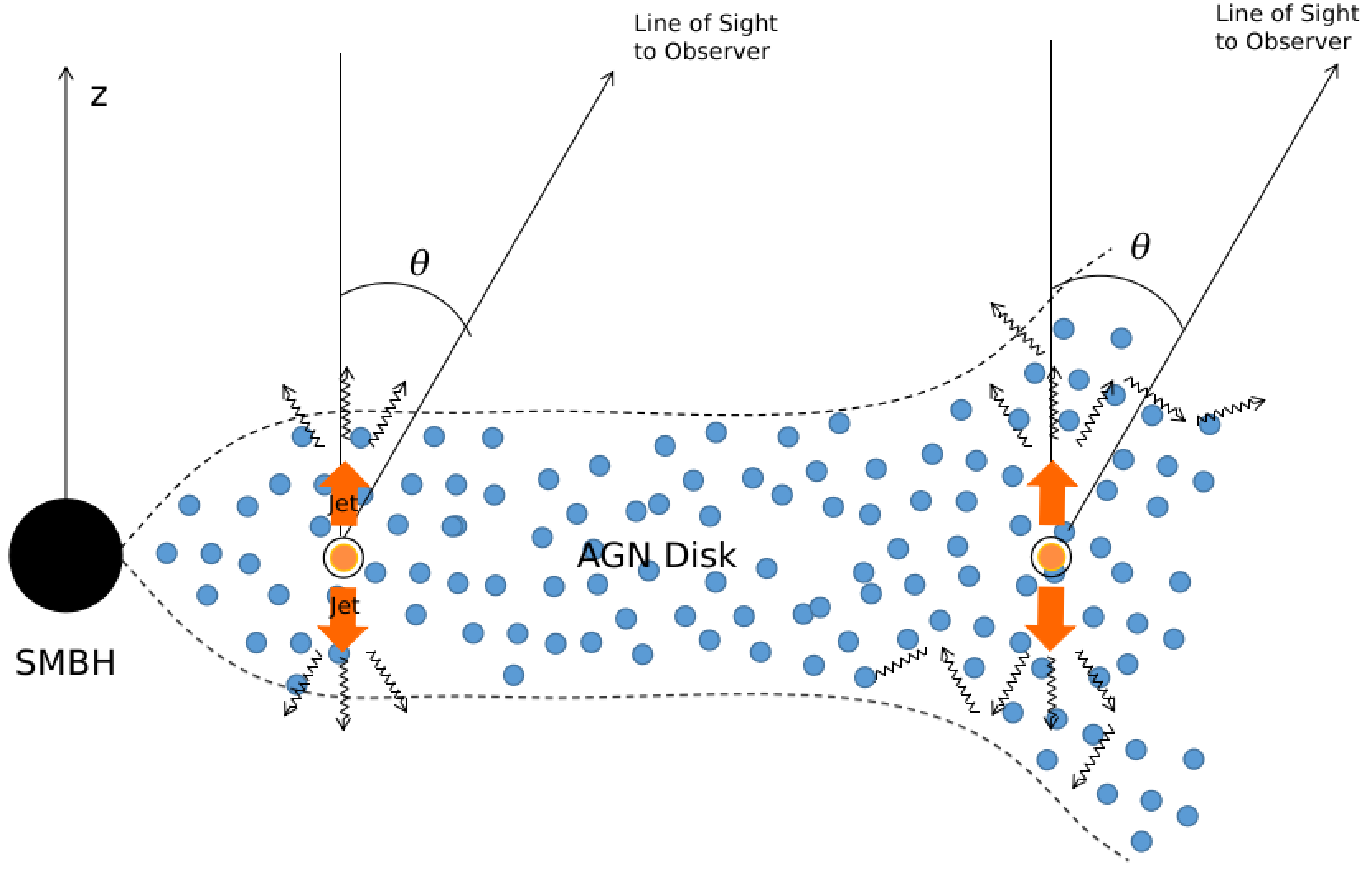}
\caption{Schematics of an afterglow source emerging from an AGN disk. The source emission is produced within a jet, whose direction is taken to be perpendicular to the direction of the plane of the disk.}
\label{fig:disk}
\end{figure}

\begin{table*}\label{tab:model-parameters}
\resizebox{\textwidth}{!}{
\begin{tabular}{ |c|c|c|c|c|c|c|c|c|} 
\hline
 Model name& $M_{\rm SMBH}$ [$M_\odot$]& r [$r_g$] & $R_{\rm ES}$ [cm] & $R_{\rm NR}$ [cm] & $R_{\rm ph}$ [cm] & $h$ [cm] & $n_0$ [$\rm cm^{-3}$] & $\bar{N}$ \\
GRB1 & $10^7$ & 700& $9.7\times10^{13}$  & $2.8\times10^{18}$ & $9.7\times10^{13}$& $5.7\times10^{13}$ & $2.3\times10^{11}$&{219}\\
GRB2 & $10^7$ & $2\times10^4$& $4.3\times10^{13}$ & $2.4\times10^{18}$ & $9.6\times10^{13}$& $3.6\times10^{13}$ & $4.1\times10^{12}$&{68644}\\
GRB3 & $10^7$& $2\times10^5$ & $1.4\times10^{14}$ & $9.2\times10^{14}$ & $3.4\times10^{15}$& $1.6\times10^{15}$ & $2.2\times10^{10}$&{2472}\\
GRB4 & $10^7$& $10^6$ & $4.6\times10^{14}$ & $4.2\times10^{15}$ & $9.8\times10^{16}$& $5.8\times10^{16}$ & $2.1\times10^{8}$&{188}\\
GRB5 & $2\times10^6$ & 700& $7.1\times10^{15}$  & $2.5\times10^{18}$ & $8.8\times10^{12}$& $1.1\times10^{13}$ & $2.3\times10^{11}$&{1.8}\\
GRB6 & $2\times10^6$ & $2\times10^4$& $1.6\times10^{15}$ & $2.5\times10^{18}$ & $1.4\times10^{13}$& $7.2\times10^{12}$ & $4.1\times10^{12}$&{1444}\\
GRB7 & $2\times10^6$& $2\times10^5$ & $1.4\times10^{14}$ & $2.3\times10^{18}$ & $4.5\times10^{14}$& $3.3\times10^{14}$ & $2.2\times10^{10}$&{44}\\
GRB8 & $2\times10^6$& $10^6$ & $4.6\times10^{14}$ & $4.2\times10^{15}$ & $8.3\times10^{15}$& $1.2\times10^{16}$ & $2.1\times10^{8}$&{1.3}\\
\hline
\end{tabular}}
\caption{Parameters of the GRB afterglow model and the surrounding medium in an AGN disk. The variable $\bar{N}=(R_{\rm ph}n_0\sigma_{\rm T})^{2}$ is the averaged number of scatterings before the photons escape the photosphere of the disk.}
\end{table*}
To propagate photons, for each individual photon with current position ($x$,$y$,$z$) and velocity ($v_x$,$v_y$,$v_z$), we first calculate the local mean free path $\lambda(r,z)$ of the photon based on the local medium number density, { e.g. a TQM disk model described in Section~\ref{sec:disk}, }  
\begin{equation}
    \lambda(r,z) = \frac{m_p}{\rho_{\rm disk}(r,z)\,\sigma_T}\,,
\end{equation}
where $m_p$ is the proton mass and $\sigma_T$ is the Thompson cross section.
{ Then a random travel distance $dr$ drawn from the exponential probability density function, }
\begin{equation}
    P(dr) =\frac{1}{\lambda(r,z)} e^{-dr/\lambda(r,z)}\,,
\end{equation}
{  is assigned to the propagating photon with isotropic orientation, }i.e. in polar coordinates $\Theta, \Phi$, $\cos{\Theta}$ is uniformly distributed in the interval $[-1,1]$, and $\Phi$ is uniformly distributed in the interval $[0,2\pi]$. After updating the velocity and position of the photon, the lab time of the photon will advance with $dt = dr/c$. Because we assume the scatterings are elastic, the frequency of photons will keep constant during the propagation. 

{ We iterate all the generated photons and keep propagating them until they all escape the photosphere of the dense medium, where the photospheric radius of the  medium (here, the AGN disk) is calculated as}
\begin{equation}
    \tau = \int_{R_{\rm ph, disk}}^\infty \rho_{\rm disk}(x)/m_p \sigma_T dx = 1
\end{equation}
where $x$ is the direction of the line of sight to the observer. { At the end of the simulation, when all the photons have escaped the photosphere of the dense medium, we record the last position, velocity and time of the photons and use this information to calculate the diffused afterglow light curves.\\
{ 
The light curve as a function of the viewing angle along the line of sight to the observer is obtained by collecting the escaped photons that have a velocity vector oriented within $\pm$ half a degree from the direction of the line of sight.}

{ To obtain a sufficient photon statistics within the small solid angle, the $10^6$ photons from the input light curve can be injected multiple times, since they become independent after the first scattering. The process is therefore repeated until a desired number of  photons is collected to generate the output light curve. Due to the nature of solid angles, lines of sight perpendicular to the disk are more challenging to model and require the propagation of more photons with respect to lines of sight that are off-axis.}
Finally, the angle-dependent light curve luminosity at time $t$ is  calculated by adding up the total energy of the weighted photons that arrive to the observer within the time window [$t-dt/2,t+dt/2$] (i.e. the surface of 'equal-arrival time'), divided by   $dt$. When the input photons are re-injected multiple times, a normalization factor $10^6/N_{\rm tot}$ is multiplied to the luminosity in the last step. 
}

\begin{figure*}
\includegraphics[width=2\columnwidth]{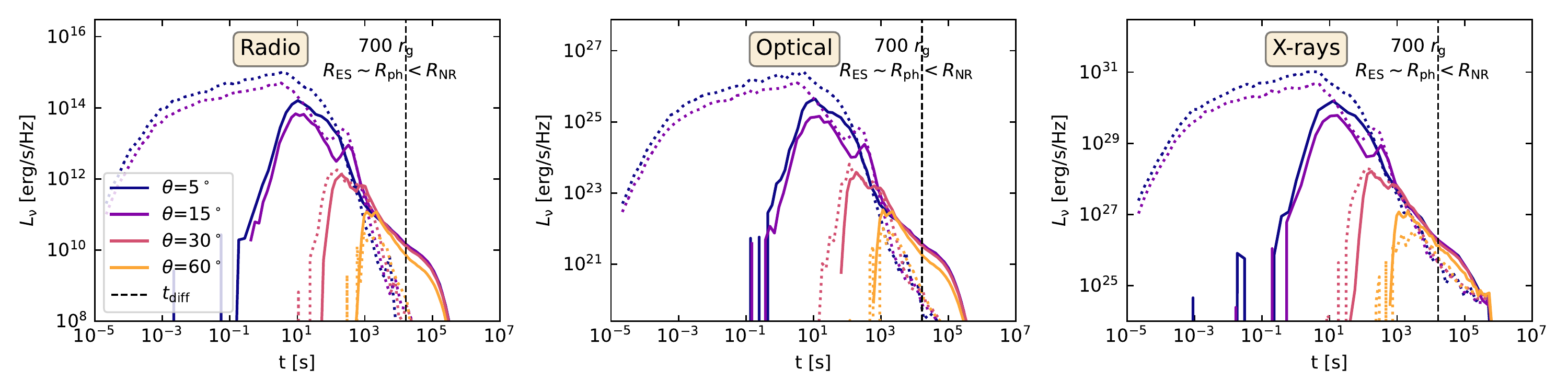}\\
\includegraphics[width=2\columnwidth]{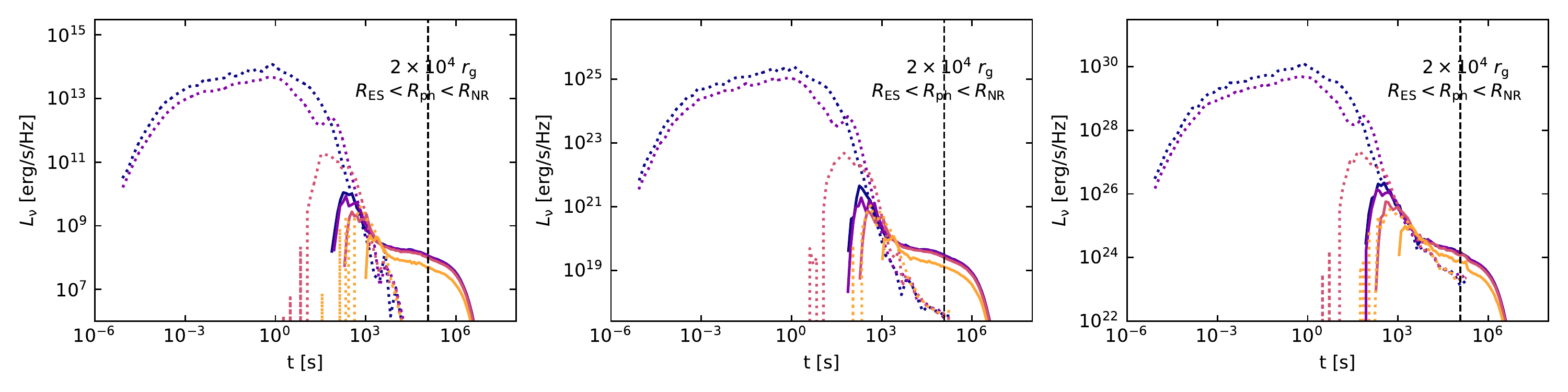}\\
\includegraphics[width=2\columnwidth]{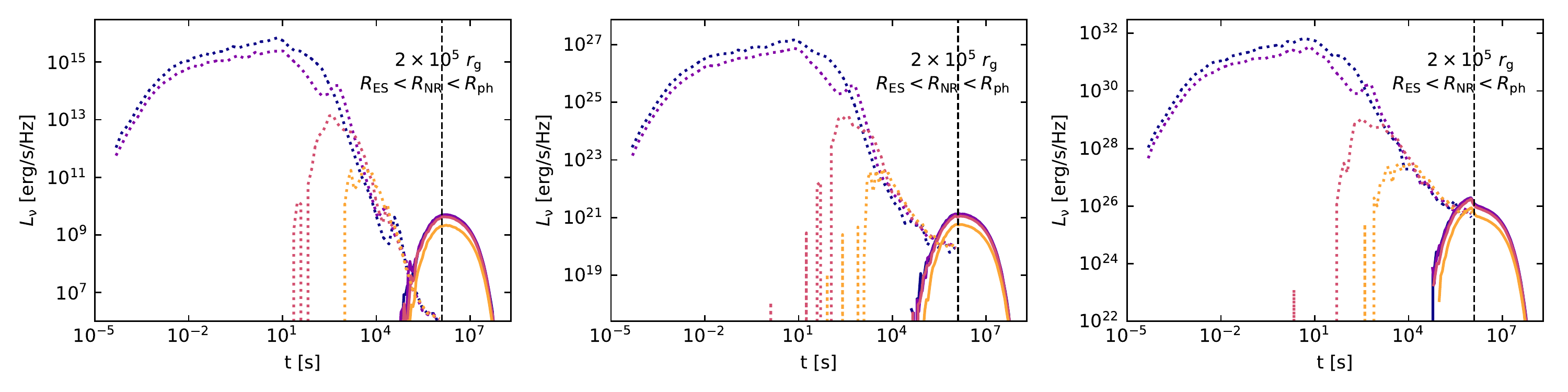}\\
\includegraphics[width=2\columnwidth]{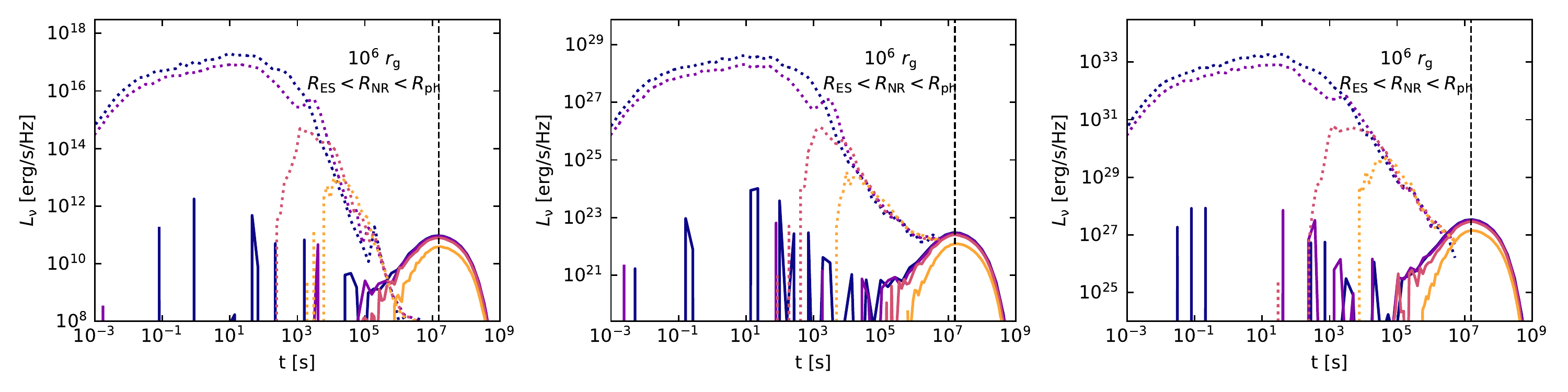}\\
\caption{Emerging afterglow light curves (for a range of viewing angles $\theta$ with respect to the jet axis)
of GRBs housed in the plane of AGN disks, at different radial locations from the central $10^7M_\odot$ SMBH, from $700~r_g$ in the top panel, to the outer region 
of $2\times10^6~r_g$ in the bottom panel. As a reference, we also plot the input light curves with dotted lines. Each chosen location is characterized by a specific value of the medium density and scale height, which influence both the input spectrum and the diffused one. For each case, we also indicate the relative magnitudes of the external shock radius (at the time the external shock forms), the non-relativistic radius, and the disk photospheric radius. 
 The light curves in the left, middle and right panels are calculated at frequencies of $10^{9}$~Hz (Radio), $5\times 10^{14}$~Hz (Optical) and $10^{17}$~Hz (X-rays), respectively. The vertical dashed line indicates the diffusion timescale (cfr. Eq.~\ref{eq:tdiff} with $H=R$). }
\label{fig:diffused-lcM1e7}
\end{figure*}

\section{Results}

\subsection{Toy model of an isotropic source in a spherical region}

As an illustrative example of how a very dense medium changes the shape of the light curve due to diffusion   {we  build  a toy model in which the source is modeled as a sharp pulse with total energy $10^{53}$~erg at some initial time  $t=10$~sec }, with isotropic emission.  It
is assumed to be located at the centre of a spherical region of uniform medium density
 and radius $10^{16}$~cm. 
 {  A graphical representation of the setup is provided in Figure~\ref{fig:toy-schematics}.}
 We perform photon scattering experiments with different number densities of the medium in a relevant range between ($10^7- 10^{10}$)~cm$^{-3}$. The  emergent light curve   for each of the densities considered 
  is shown in the left panel of Figure~\ref{fig:toy}, while  the 
  cumulative density function for the 
  number of electron-photon scatterings during the propagation of each photon is correspondingly shown in the middle panel of the same figure. Note that in the lower density cases there is a fraction of photons which emerges without being scattered (i.e. {$N+1=1$} in the middle panel of the figure). These photons are the ones which emerge in correspondence of the time of the input pulse at $t=10$~sec, thus producing the spike at that time which is evident in the left panel of the figure. { The horizontal dashed lines indicate the transmittance of the medium that matches the fraction of unscattered photons for each case. The right panel of Figure~\ref{fig:toy} shows the cumulative distribution of the angle $\theta_{\rm v, r}$ between the velocity vector and the location vector at the last scattering. For a low density medium where the mean free path of the photons is larger than the size of the medium, most of photons remain unscattered, hence traveling in the radial direction with $\cos\theta_{\rm v, r}=1$. As the optical depth increases, an increasingly larger number of photons is scattered,  thus exiting the medium in various directions. However, as the optical depth becomes large than one, geometry beaming effects due to the geometry of the surface come into play. Thus we can see more photons with small $\theta_{\rm v, r}$.
}
\begin{figure*}

\includegraphics[width=2\columnwidth]{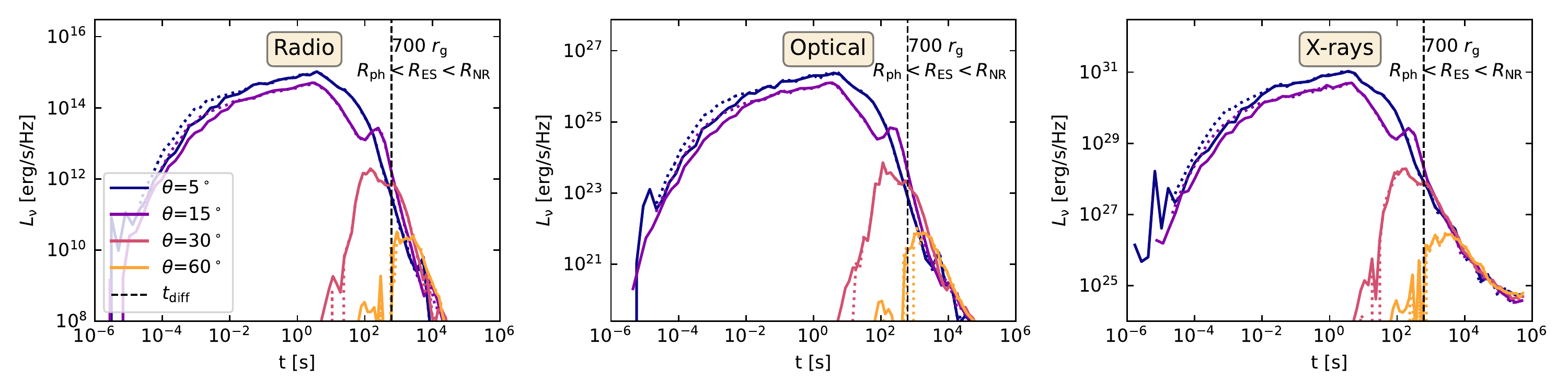}\\
\includegraphics[width=2\columnwidth]{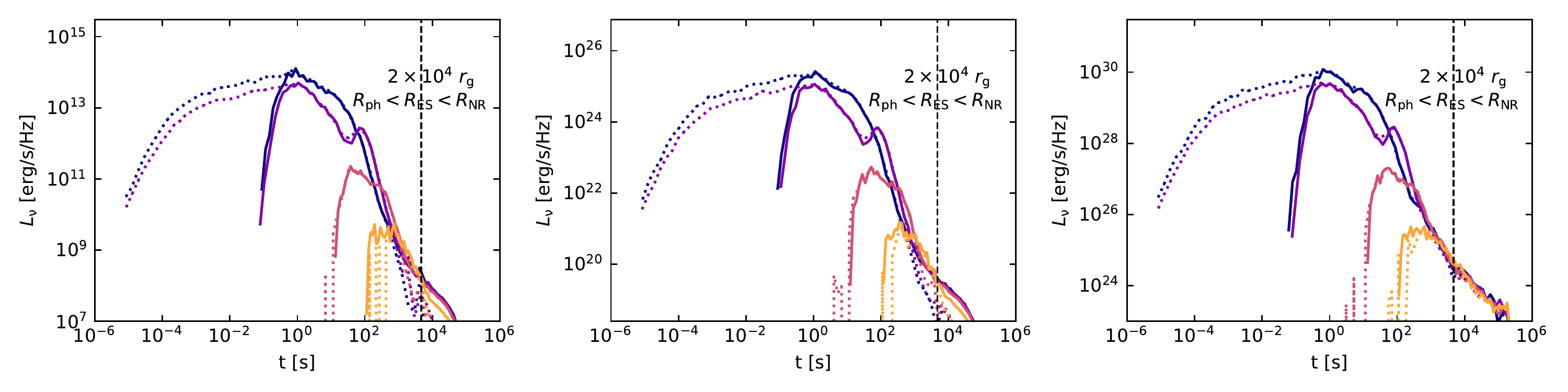}\\
\includegraphics[width=2\columnwidth]{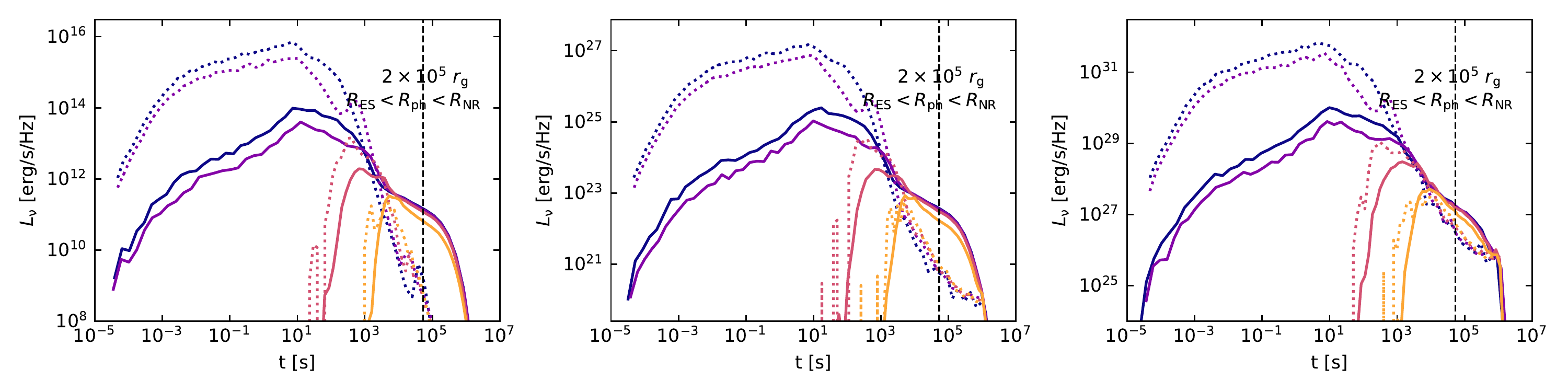}\\
\includegraphics[width=2\columnwidth]{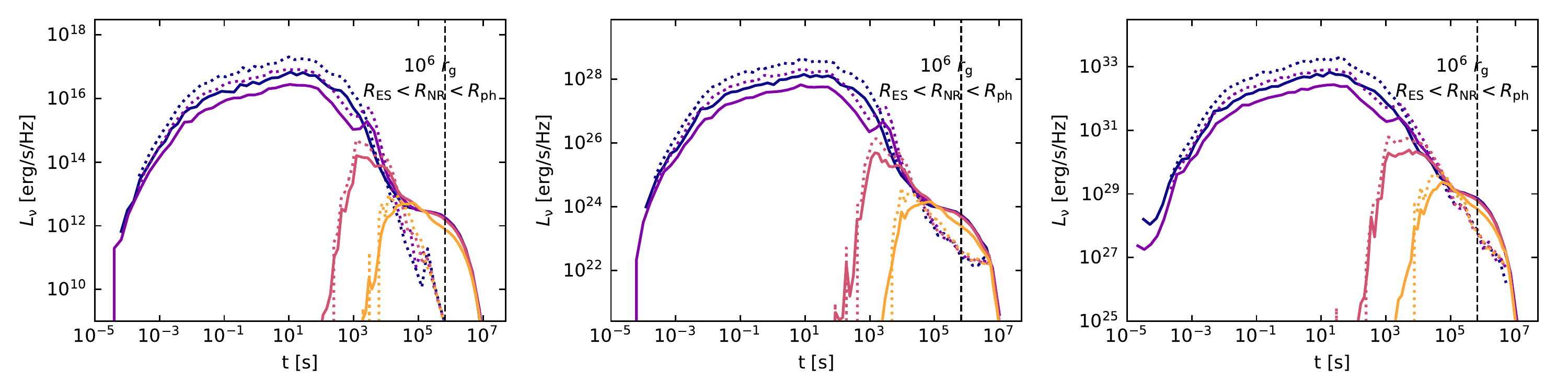}\\
\caption{Same as in Fig.~\ref{fig:diffused-lcM1e7}, but for a disk around a SMBH of $2\times 10^6 M_\odot$. }
\label{fig:diffused-lcM2e6}
\end{figure*}

The toy model already illustrates the main features of a diffused afterglow: it emerges spread out in time on the diffusive timescale 
\begin{equation}
    t_{\rm diff} \approx \frac{R^2 \rho_0 \sigma_T}{m_p c} \approx 2.2\times 10^7
    \left(\frac{R}{10^{16}~{\rm cm}}\right)^2
    \left(\frac{n_0}{10^{10}~{\rm cm}^{-3}}\right)~{\rm sec}\,,
\label{eq:tdiff}
\end{equation}
where $R$ is the size of the region and $n_0=\rho_0/m_p$ its density. 
Correspondingly, if $E$
is the {total  energy of the input light curve, the luminosity of the emergent light curve is on the order of $\sim \frac{E} {t_{\rm diff}}\frac{\Omega}{4\pi}$, where $\Omega$ is {  the jet solid angle}.}

\subsection{Jetted sources from AGN disks}
As described in Section~\ref{sec:GRB-afterglow}, we generate a model of jetted GRB and its afterglow with an asymptotic Lorentz factor $\Gamma_0=300$, outflow energy $E_{\rm iso}=10^{53}$ erg, burst engine duration $T_{\rm eng}=20$ s, electron fraction $Y_{\rm e}=1$, 
typical timescale of emission between two shells $\Delta t = 0.1$~s, 
and jet opening angle $\Delta \theta = 15^\circ$. For GRBs produced in AGN disks, there are a few important radii \citep{Perna2021} whose relative location determines the evolution of the fireball and the observability of the afterglow: 
the external shock radius $R_{\rm ES}$, where the fireball dissipates its energy and an afterglow is formed, the non-relativistic radius $R_{\rm NR}$, where the mass swiped up is large enough to lead the blastwave to decelerate to non-relativistic speed, and the disk photospheric radius $R_{\rm ph}$, where the photons' mean free path for Thompson scattering becomes infinite.

For fireballs with $R_{\rm ES} < R_{\rm ph} < R_{\rm NR}$, the early radiation of the afterglow will be diluted until the external shock exits the photosphere of the disk. For fireballs with $R_{\rm ES} < R_{\rm NR} < R_{\rm ph}$, the afterglow emission will be fully isotropized and time-diluted. 

We modeled jetted GRB afterglows at different locations of an AGN disk around a $10^7$ $M_\odot$ and a $2\times10^6$ $M_\odot$ supermassive black hole, especially focusing on situations in which the photosphere of the disk is either larger or smaller than the non-relativistic radius of the GRB afterglow. 
Table~1 summarizes the parameters of the GRB afterglow and the AGN disk at the initial location of the source (recall
that the jet, while emitting, is also advancing within the AGN disk).

The emerging afterglow light curves for each of the models of an AGN disk with SMBH of $10^7$ $M_\odot$ in Table~1 are displayed in  Fig.~\ref{fig:diffused-lcM1e7} for the three representative bands: Radio at 9~GHz, Optical at $5\times 10^{14}$~Hz, and X-rays at $10^{17}$~Hz.
For each case, we compute the emerging radiation for four viewing angles (between $5^\circ$ and $60^\circ$) with respect to the jet axis which, for simplicity, is assumed to coincide with the direction perpendicular to the AGN disk
as illustrated in Fig.~\ref{fig:disk}.

We begin by noticing some features which are 
primarily the result of the peculiarity of the emission spectrum in very dense media, as discussed in Sec.2.2 and illustrated in 
Fig.~\ref{fig:lightcurves}.
More specifically, as a general feature (i.e. at all GRB locations), 
the radio emission is extremely weak,
as a result of self-absorption during the process of the production of the radiation itself. 
The optical is also suppressed at emission by about 6-7 orders of magnitude compared to a similar source in an interstellar medium environment. 
Last, the X-rays emission is the least affected by the very high density medium.

For GRB sources located at 700 $r_{\rm g}$,
where the external shock forms close to the disk photosphere, only
 the very early afterglow radiation is diffused and hence suppressed. However
 the external shock very quickly reaches the photosphere, after about 1~sec, and from that moment onward the photons at the shock are generated outside of the photosphere of the disk. Hence the emergent 
  light curves are similar to the undiffused ones until the early diffused photons emerge on the diffusion timescale at $t\sim t_{\rm diff}$, adding to the tail of the light curves. 
  Because only the very early afterglow radiation is diffused by the medium,  the dependence of the emergent light curves on the viewing angle (angle that the observer makes with the jet axis)
   is almost identical to that of the undiffused afterglow light curves, where observers at larger viewing angles cannot see the early radiation due to its strong relativistic beaming.

If the GRB occurs in the region of the disk where the photosphere of the disk is larger than the external shock radius, more photons from the afterglow will be diluted until the external shock exits the photosphere of the disk. With $\Gamma_0=300$, the external shock takes roughly $\sim 10^2-10^3$~s to exit the photosphere of the disk. Hence we can see that the light curves in the second row of Figure~\ref{fig:diffused-lcM1e7} start roughly at $10^3$~s. More diffused photons emerge later on the timescale $t_{\rm diff}$, leading to a flatter tail of the light curves.

For GRBs in a region of the AGN disk where the photosphere of the disk is larger than the non-relativistic radius as shown in the third row of Figure~\ref{fig:diffused-lcM1e7}, the afterglow radiation will be  diffused on all timescales, and its intensity suppressed 
by a factor on the order of {$\frac{t_{\rm diff}}{T_{\rm lc}}\frac{\Omega}{4\pi}$,} where $T_{\rm lc}$ is the duration of the undiffused transient and $\Omega$ is the {  jet solid angle}.
In this case, viewing angle effects due to the relativistic beaming of the radiation will also be completely washed out due to the (almost) isotropic Thompson scatterings.   

We further studied the emergence of GRB afterglows from a location of $10^6$ $r_g$ in the disk, where the number density of the medium is $n_0\sim 10^8$ cm$^{-3}$. Here the 
suppression of the input emission due to
self-absorption is reduced compared to the
inner disk regions discussed above, hence
producing a brighter afterglow than in those regions, as shown in Figure~\ref{fig:diffused-lcM1e7}.  The shape of the emergent light curves is similar to that for the case $r=2\times10^5 r_g$. Fully diffused light curves with luminosity reduced by a factor of { $\frac{t_{\rm diff}}{T_{\rm lc}}\frac{\Omega}{4\pi}$ }emerge on a timescale on the order of  $t_{\rm diff}$.

\begin{figure}
\includegraphics[width=\columnwidth]{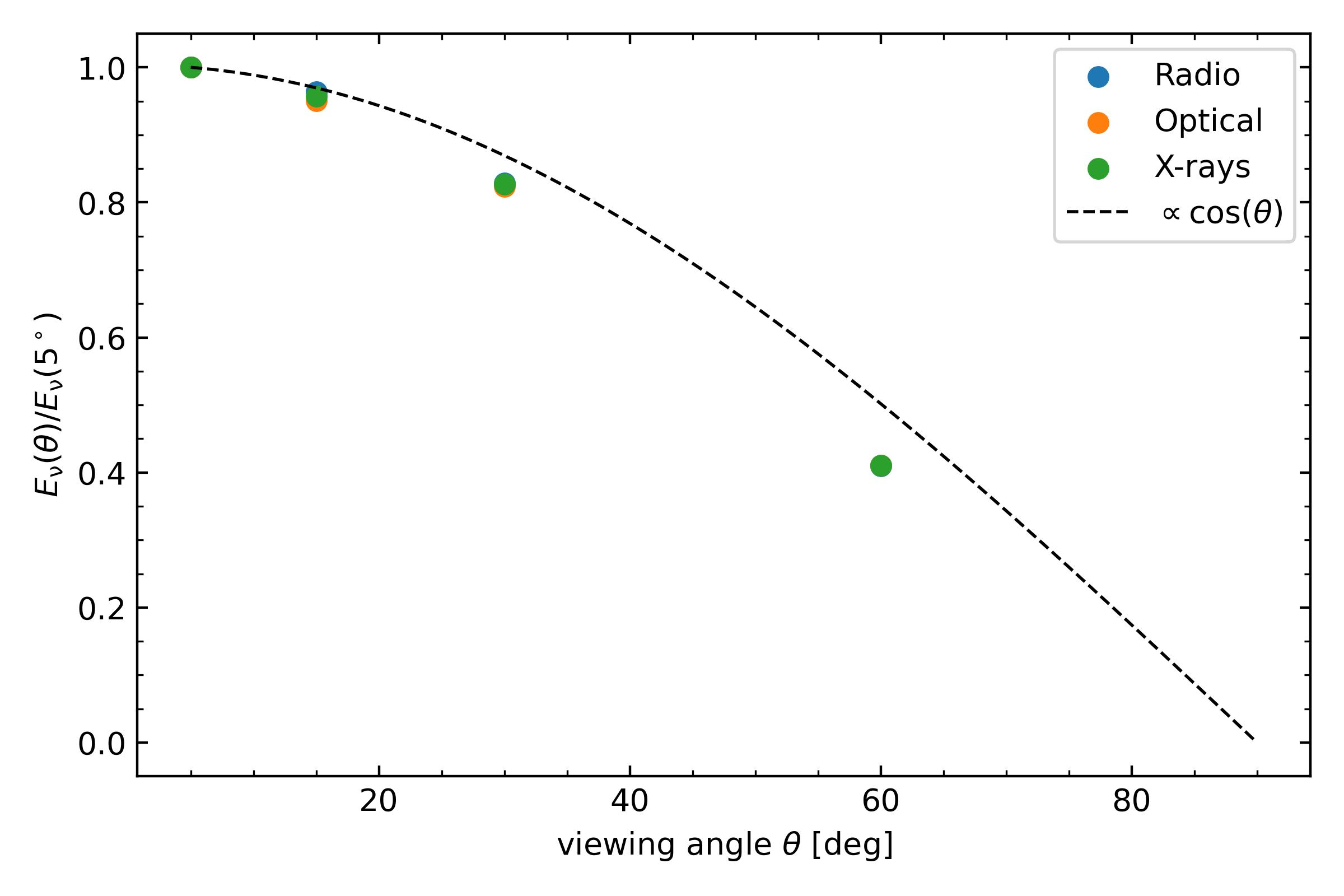}\\
\caption{Fluence of fully diffused GRB afterglows as a function of the viewing angle in the three analyzed bands. Values are normalized to the $5^\circ$ off-axis observer. The data are specifically for model GRB3. The theoretical limit for $\tau\to\infty$ is overlaid with a dashed line.}
\label{fig:bol-lum}
\end{figure}
{Figure~\ref{fig:bol-lum} shows the angular dependence of the fluence of fully diffused afterglows for the specific model in the third row of Figure~\ref{fig:diffused-lcM1e7} (GRB3), as a function of the viewing angle $\theta$. Even though the photons velocity vectors are largely isotropized inside the disk after many scatterings, the radiation emerging from the disk surface is beamed preferentially along the normal to the surface. This is due to the fact that a photon propagating along the normal direction needs to travel a shorter path to reach the surface and therefore has a lower probability of being scattered again before emerging. If the emitting surface were infinite and the radiation fully isotropized ($\tau\to\infty$), then the beaming would follow a $\cos\theta$ beaming profile, where theta is the polar angle away from the normal. Our data show a slightly higher beaming, likely related to the fact that the radiation is not completely made isotropic by scattering.}

As a second case study, we considered an AGN disk around a SMBH of mass $2\times10^6$ $M_\odot$ (models 5 { through} 8 in Table~1), and studied the emergence of GRB afterglow transients at the same locations as for the higher mass case discussed above. For the smaller SMBH case,
 the accretion disk is geometrically and optically thinner, thus yielding a smaller size of the disk photosphere. Afterglows in this thinner disk are hence less diffused,
 as shown in Fig.~\ref{fig:diffused-lcM2e6}.
More specifically, inspection of this figure
 shows that,  for sources located at $700~r_g$, the light curves are almost identical to the undiffused ones due to a relatively small size of the photosphere of the disk. As the photosphere of the disk becomes larger at $2\times10^4$ $r_g$, a small fraction of the early photons are diffused. The later parts of the light curves remain similar to those of the input afterglows until the diffused photons emerge on the diffusion timescale. For afterglows in the outer region of the disk, the number density of the electrons is relatively smaller, and the mean free path of the photons is thus relatively larger. The disk scale height is smaller than it is at the same location of the disk in the higher ($10^7M_\odot$) SMBH case, where as discussed most photons are scattered.
 For the $2\times 10^6M_\odot$ SMBH disk on the other hand, even if the photosphere of the disk is larger than the non-relativistic radius, only a small fraction of photons will be scattered, thus resulting in light curves similar to the undiffused ones.

\section{Discussion and Conclusions}

In this work, building on our first investigation of GRBs emerging from AGN disks, we have computed the detailed afterglow luminosity, combining a generalized source luminosity in the very high density regime, with a Monte Carlo radiative transfer code for the propagation of the source photons within the AGN disk, until their emergence at the disk photosphere.

As expected, we have found that the transients are dimmer but last longer, and we have quantified the emerging shape of the light curve, characterized by a sharp rise and fall, and a smooth, long-lived quasi plateau, with peak luminosity being reached after a few diffusive timescales.
While their typical durations, on the order of hundreds of days, makes them ideal candidates for detection by transient surveys such as eRosita, ZTF and VRO, however their identification requires their emission to emerge above that of the AGN itself.

Most AGNs have bolometric luminosities $10^{43-47}$~erg~s$^{-1}$ spread across a wide spectrum but with a large component in the UV/optical
(e.g. \citealt{Woo2002}). There  appears to be no strong correlation with the SMBH mass, and AGNs with comparable mass for the central SMBH may have bolometric luminosities differing by several orders of magnitude. 
Therefore, in general terms, the search for diffused GRB afterglow transients
will be more effective among the lower-luminosity AGNs.
In radio (5GHz, \citealt{Woo2002}), the luminosity varies in the $\sim 10^{27}-10^{35}$~erg/s/Hz range, with the majority of sources being either in the high-luminosity range (radio loud) or the low-luminosity one (radio quiet). 

In X-rays, the AGN luminosity function is characterized by a broken power law (i.e. \citealt{Gilli2007}). Measurements of the 2-10~keV luminosity in the low redshift Universe \citep{Ueda2003} show that, while there are sources as bright as $\sim 10^{46-47}$~erg/s, these are however suppressed in number by about $\sim 7$ orders of magnitude with respect to the number of sources with luminosities in the $10^{41-42}$~erg/s range.

Below this range, it exists
an interesting subclass of AGNs
called 'low-luminosities AGNs', which have bolometric luminosities in the $\sim 10^{39-41}$~erg~s$^{-1}$ (e.g. \citealt{Maoz2007}); 
these would be the most useful targets for the detection of GRB afterglow transients.  However, as evident from Figs.~\ref{fig:diffused-lcM1e7} and \ref{fig:diffused-lcM2e6},
detection prospects are significantly weighed towards higher energies.
Given the strong suppression at low frequencies of the input spectrum, the radio afterglow is undetectable even in the most optimistic case of transients located in the external regions of the disk where the densities are less extreme and diffusion is less severe. 

Detection in the optical bands is significantly more promising, albeit only for smaller SMBH disks, and with preferences in some specific regions of the disk. For an AGN disk around a SMBH of $10^7 M_\odot$, on-axis transients (i.e. within $15^\circ$ for the jet model adopted here)
from locations at around $700~r_g$ would have optical luminosities around $\sim 10^{39}-10^{40}$~erg~s$^{-1}$, and hence detectable in low-luminosity AGNs. For an AGN with a SMBH of $2\times 10^6M_\odot$, an optical GRB afterglow seen on axis would have a luminosity of $\sim 10^{41}$~erg~s$^{-1}$   at the same location of $700 r_g$, and of {$5 \times 10^{42}$}~erg~s$^{-1}$ at outer radii of $10^6 r_g$. Both these locations are interesting from an astrophysical perspective. At closer radii to the SMBH,  there are regions in which migration traps are formed \citep{Bellovary2016}, leading to strong dynamical interactions between compact objects and hence formation of binary neutron stars and neutron star-black hole binaries, which
yield short GRBs (for neutron star-black hole mergers it further depends on the mass ratio). At larger radii, AGN disks are Toomre unstable \citep{Levin2003}, which facilitates the formation of stars
and hence long GRBs \citep{Jermyn2021}. 

In the X-rays on the other hand, afterglow transients will generally emerge well above the AGN disk luminosity, with emission at 10~keV (from on-axis viewing angles) ranging from a minimum of $\sim 10^{43}$~erg~s$^{-1}$
to as high as $\sim 10^{50}$~erg~s$^{-1}$ among all the locations and SMBH masses we considered. Hence rebrightenings of AGNs in the X-rays, over timescales of hundreds to thousands of seconds, constitute very strong candidates of GRB afterglows. If the rebrightening happens simultaneously also at longer wavelengths, albeit with smaller fluence, it would further signal that the transient is a candidate GRB afterglow.

Being able to recognize, and hence calibrate, the population of long and short GRBs in AGN disks will have important implications for learning about the presence of compact objects in these disks and their interaction, which is of special relevance for the AGN channel of the LIGO/Virgo observations \citep{McKernan2020,Wang2021}, as well as for the evolution of stars 
in these special AGN environments \citep{Cantiello2021,Jermyn2021,Dittmann2021,Jermyn2022}.

\section*{Acknowledgements}
 DL acknowledges support from NSF grant AST-1907955.
RP and YW acknowledge support by NSF award AST-2006839.

\section*{Data Availability}

 All the data produced for this project will be provided upon reasonable request to the authors.



\bibliographystyle{mnras}
\bibliography{biblio} 







\bsp	
\label{lastpage}
\end{document}